\documentclass{emulateapj}
\usepackage{natbib}
\usepackage{epsfig} 
\usepackage{graphicx}
\usepackage{subfigure}
\usepackage{float}
\usepackage{amsmath}
\usepackage{color}
\usepackage{amssymb} 
\usepackage{amsfonts}
\usepackage{units}
\usepackage{bm} 
\usepackage[colorlinks,linkcolor=blue,anchorcolor=green,citecolor=blue]{hyperref}

\shorttitle{DM variation of repeating FRB sources}
\shortauthors{Yang \& Zhang} 
\begin{document} 

\title{Dispersion Measure Variation of Repeating Fast Radio Burst Sources}

\author{Yuan-Pei Yang\altaffilmark{1,2} and Bing Zhang\altaffilmark{1,3,4}}

\affil{$^1$Kavli Institute for Astronomy and Astrophysics, Peking University, Beijing 100871, China; yypspore@gmail.com;\\
$^2$ KIAA-CAS Fellow\\
$^3$ Department of Astronomy, School of Physics, Peking University, Beijing 100871, China \\
$^4$ Department of Physics and Astronomy, University of Nevada, Las Vegas, NV 89154, USA; zhang@physics.unlv.edu\\
}

\begin{abstract}
The repeating fast radio burst (FRB) 121102 was recently localized in a dwarf galaxy at a cosmological distance. The dispersion measure (DM) derived for each burst from FRB 121102 so far has not shown significant evolution, even though an apparent increase was recently seen with newly detected VLA bursts. It is expected that more repeating FRB sources may be detected in the future. In this work, we investigate a list of possible astrophysical processes that might cause DM variation of a particular FRB source. The processes include (1) the cosmological scale effects such as Hubble expansion and large-scale structure fluctuations; (2) the FRB local effects such as gas density fluctuation, expansion of a supernova remnant, a pulsar wind nebula, and an HII region; and (3) the propagation effect due to plasma lensing. We find that the DM variations contributed by the large-scale structure are extremely small, and any observable DM variation is likely caused by the plasma local to the FRB source. Besides mechanisms that produce decreasing DM with time, we suggest that an FRB source in an expanding supernova remnant around a nearly neutral ambient medium during the deceleration (Sedov-Taylor and snowplow) phases or in a growing HII region can introduce DM increasing. Some effects (e.g. an FRB source moving in an HII region or plasma lensing) can give either positive or negative DM variations. Future observations of DM variations of FRB 121102 and other repeating FRB sources can bring important clues for the physical origin of these sources. 
\end{abstract}

\keywords{intergalactic medium --- ISM: general --- radio continuum: general}

\section{Introduction}

Fast radio bursts (FRBs) are millisecond-duration radio transients characterized by bright fluence ($\sim 1~\unit{Jy-ms}$), large dispersion measure (DM, which is $\gtrsim200~\unit{pc~cm^{-3}}$), and high all-sky rate ($\sim500-10^4~\unit{events~sky^{-1}d^{-1}}$) \citep{lor07,kea12,kea16,tho13,bur14,spi14,spi16,mas15,pet15,rav15,rav16,sch16,cha16,law16,van16,cha17,cal17,pet17,ban17}. 
So far 23 FRBs have been published, among which one case, FRB 121102, clearly shows a repeating behavior \citep{spi16,sch16,cha17,law17,sch17} and was recently localized in a star-forming galaxy at $z=0.19273$ thanks to the precise localization and multi-wavelength follow-up observations \citep{cha17,ten17,mar17}. Other non-repeating FRBs are also suggested to be of an extragalactic or even a cosmological origin, due to their all-sky distribution and the DM excess with respect to the Galactic contribution. 

For an electromagnetic wave propagating in a cold plasma, the arrival time of a pulse has a power-law dependence on frequency, i.e. $\Delta t\propto\nu^{-2}{\rm DM}$, where the dispersion measure DM is given by\footnote{Notice that for plasma from a cosmological distance, the observed DM should be corrected as ${\rm DM}=\int n_e/(1+z)dl$ \citep[e.g.][]{iok03,ino04,den14}.}
\begin{eqnarray}
{\rm DM}=\int n_e dl,
\end{eqnarray}
which is the column density of free electrons along the line of sight, where $n_e$ denotes the volume number density of free electrons any spatial point along the line of sight. For an FRB, the observed DM has three contributions, which are from Milky Way, the intergalactic medium (IGM), and the host galaxy (including the interstellar medium (ISM) and the near-source plasma), respectively. In particular, the contribution from the host galaxy plays an important role in identifying the origin of FRBs \citep[e.g.][]{ten17}. For example, a large host-galaxy DM might imply a significant contribution from the near-source plasma, such as supernova remnant (SNR), pulsar/magnetar wind nebula (PWN), or HII region \citep{yan17}.

Besides the absolute DM value, the variation of DM with time also carries important information to diagnose the properties of the medium along the propagation path. This is particularly relevant for a repeating FRB source such as FRB 121102. So far, only FRB 121102 is confirmed to repeat. Another candidate as suggested by \cite{pir17} is that 
FRB 110220 and FRB 140514, which are in the same 14.4 arcmin beam and 9 arcmin apart with the latter having a smaller DM than the former, might originate from the same neutron star embedded within a SNR. They suggested that a detection of a third burst in the direction with an even smaller DM would confirm the hypothesis and lead to a derivation of the age of the neutron star.  \cite{kat17} suggested that one may test the FRB model of massive star core collapse without a supernova using a long-term observation of the DM variation of a repeating FRB source. Observationally, the host galaxy of FRB 121102 reveals a bright star forming region and a compact radio source around the location of the FRB source \citep{cha17,mar17,ten17,kok17,bas17}. It may imply that there are dense-plasma regions in the host galaxy, one of which may contribute to the observed DM and may even work as a plasma lens \citep{cor17}.

Over a hundred bursts have been detected from the FRB 121102 source \citep{spi16,sch16,cha17,law17,sch17}. 
Recently reported nine VLA bursts have a mean apparent DM of $567.8\pm 0.1~\unit{pc~cm^{-3}}$ \citep{law17}, which is significantly larger than the previously reported long-term average of $560.5~\unit{pc~cm^{-3}}$ detected by Arecibo \citep{spi14,spi16,sch16,cha17}. 
For a given frequency band ($1.4~\unit{GHz}$ for Arecibo; $3~\unit{GHz}$ for VLA), the DM does not show a significant evolution tendency during long-term observations, e.g. four years for Arecibo; one month for VLA \citep{law17}. \cite{law17} suggested that the difference in different apparent DM values from VLA and Arecibo may be due to some un-modeled, frequency-dependent, source-intrinsic effects (e.g. different intrinsic pulse profiles in different frequency bands), so that it might not be a secular intrinsic DM variation from the source.

In this paper, we investigate the possible physical mechanisms that may cause DM variation of a repeating FRB source, and suggest that the detection of intrinsic DM variation can be used to diagnose the local environment of repeating FRBs.

\section{Physical mechanisms for DM variation}

In general, DM variation can be attributed to the density fluctuation of free electrons along the line of sight, or the distance variation between the FRB source and the observer. In the following, we discuss a list of possible DM variation mechanisms ranging from cosmological effects to local effects.

\subsection{Hubble expansion}

The largest scale effect of the DM variation may be due to the expansion of the universe. The mean DM caused by the IGM is given by \citep[e.g.][]{den14,yan16}\footnote{\cite{iok03} presented an expression of DM as a function of $z$ for the first time, but his value is larger by a factor of $\sim 1.38$ due to two simplifications: first, he assumed that all baryons are in the IGM, so that a factor $f_{\rm IGM} \sim 0.83$ \citep{fuk98,shu12} is missing; second, he did not consider the He content of the universe, so that a factor $f_e \sim 7/8$ is missing.}
\begin{eqnarray}
{\rm{DM_{IGM}}}=\frac{3cH_0\Omega_{\rm{b}}f_{\rm IGM}}{8\pi Gm_p}\int_0^z\frac{f_e(z^\prime)(1+z^\prime)}{\sqrt{\Omega_{\rm m}(1+z^\prime)^3+\Omega_\Lambda}}dz^\prime,\nonumber\\\label{dmigm}
\end{eqnarray}
where $H_0$ is the current Hubble constant, $\Omega_{\rm b}$, $\Omega_{\rm m}$ and $\Omega_\Lambda$ are the current mass fractions of baryon, matter and dark energy, respectively, $f_{\rm{IGM}}$ is the fraction of baryon mass in the intergalactic medium which adopted as $\sim 0.83$ \citep{fuk98,shu12}, and $f_e=n_{e,\rm{free}}/(n_p+n_n)$ is the number ratio between free electrons and baryons (including proton and neutron) in IGM. For nearby FRBs at $z<3$, since both hydrogen and helium are fully ionized, one has $n_{e,\rm{free}}\simeq n_e=n_p$. On average, in the universe $n_p:n_n\simeq7:1$, one therefore has $f_e\simeq 7/8$. According to Eq.(\ref{dmigm}) and 
the redshift-time relation, e.g., $dz/dt=-H_0(1+z)\sqrt{\Omega_{\rm m}(1+z)^3+\Omega_{\Lambda}}$, the DM variation contributed by the Hubble expansion is therefore given by
\begin{eqnarray}
\frac{d{\rm DM_{IGM}}}{dt}&=&\frac{d\rm{DM_{IGM}}}{dz}\frac{dz}{dt}=-\frac{3cH_0^2\Omega_{\rm{b}}f_{\rm IGM}f_e}{8\pi Gm_p}(1+z)^2\nonumber\\
&\simeq&-5.6\times10^{-8}(1+z)^2~\unit{pc~cm^{-3}yr^{-1}}.\label{vdmigm}
\end{eqnarray}
Here, we have adopted the flat ${\rm \Lambda CDM}$ parameters derived from the \emph{Planck} data: $H_0=67.7~\unit{km~s^{-1}Mpc^{-1}},~\Omega_{\rm m}=0.31,~\Omega_\Lambda=0.69$, and $\Omega_{\rm b}=0.049$ \citep{pla15}. 
Therefore, the DM variation due to Hubble expansion is very small. Interestingly, we note that Eq.(\ref{vdmigm}) does not depend on  the cosmological parameters $(\Omega_{\rm m},\Omega_\Lambda)$.

\subsection{Large scale structure (LSS) fluctuations}

\subsubsection{LSS gravitational potential fluctuation}

We consider the DM variation resulting from the fluctuations of the gravitational potential of the LSS. According to general relativity, the Shapiro delay time in a gravitational potential $U(r,t)$ is given by\footnote{Note that this equation is based on the Schwarzschild metric, which we assume that the gravitational potential of LSS approximately satisfies.}
\begin{eqnarray}
\delta t_{\rm gra}=-\frac{2}{c^3}\int_{r_o}^{r_e} U(r,t)dr, \label{sha}
\end{eqnarray}
where the integration is along the path of the photon emitted at $r_e$ and received at $r_o$. 
As shown in Eq.(\ref{sha}), for a localized FRB, a fluctuation of the LSS gravitational potential along line of sight would cause a slight variation of the optical path, leading to a DM variation. 

The gravitational potential of LSS can be calculated via the observed peculiar velocities, $v_p$, of galaxies \citep[e.g.][]{nus16}. 
The gravitational acceleration produced by the mass distribution fluctuations may be approximated as $g\sim v_p/t_{\rm acc}\sim v_pH_0$, where $t_{\rm acc}$ is the acceleration timescale caused by the LSS, which corresponds to the only possible cosmological timescale, e.g. $H_0^{-1}$.  
For a sphere with a radius $R\sim100~\unit{Mpc}$ around us, the typical bulk peculiar velocity is $v_p\sim300~\unit{km~s^{-1}}$. 
Thus, the LSS gravitational potential may be given by 
$U_{\rm{LSS}}\sim gR\sim v_p RH_0\sim2.2\times10^{16}~\unit{cm^2s^{-2}}$ \citep{nus16}, which is of the same order of magnitude as the potential fluctuations inferred from the fluctuations of the cosmic microwave background \citep{ben94}. 
The difference of the Shapiro delay time during the observed time $T$ may be given by
\begin{eqnarray}
|\Delta t| &\simeq&\left|\frac{\partial(\delta t_{\rm gra})}{\partial t}\right|T\simeq\frac{2T}{c^3}\int_{r_o}^{r_e}\left|\frac{\partial U(r,t)}{\partial t}\right|dr\simeq\frac{2H_0}{c^3}U_{\rm{LSS}}DT\nonumber\\
&\simeq&0.1~\unit{hr} \left(\frac{U_{\rm{LSS}}}{2.2\times10^{16}~\unit{cm^2s^{-2}}}\right)\left(\frac{D}{1~\unit{Gpc}}\right)\left(\frac{T}{1~\unit{yr}}\right),\nonumber\\
\end{eqnarray}
where $D$ is the distance between an FRB source and the observer. The observed time $T$ corresponds to the time interval between the first and the second FRB detection over which the difference in DM is being tested. The evolution timescale of the LSS gravitation potential is approximately $H_0^{-1}$.
The DM variation is estimated as
\begin{eqnarray}
\left | \frac{d\rm DM_{LSS,G}}{dt} \right |&\sim& \frac{n_ec | \Delta t|}{T}\simeq3.5\times10^{-13}~\unit{pc~cm^{-3}yr^{-1}}\nonumber\\
&\times&\left(\frac{n_e}{10^{-7}~\unit{cm^{-3}}}\right)\left(\frac{U_{\rm{LSS}}}{2.2\times10^{16}~\unit{cm^2s^{-2}}}\right)\left(\frac{D}{1~\unit{Gpc}}\right).\nonumber\\
\end{eqnarray}
Here, we have assumed that the mean number density of the free electrons does not significantly change during the gravitational potential fluctuation. One can see that this effect is even less important than Hubble expansion.

\subsubsection{LSS density fluctuation}
 
Next, we consider gas density fluctuation in the LSS. 
As discussed above, for LSS, the only possible fluctuation timescale is $H_0^{-1}$. Thus, the DM variation caused by the LSS gas density fluctuation could be estimated as
\begin{eqnarray}
\left | \frac{d\rm DM_{LSS,D}}{dt} \right |&\sim& |\delta n_e| D H_0\simeq6.9\times10^{-9}~\unit{pc~cm^{-3}yr^{-1}}\nonumber\\
&\times&\left(\frac{|\delta n_e|}{10^{-7}~\unit{cm^{-3}}}\right)\left(\frac{D}{1~\unit{Gpc}}\right),
\end{eqnarray}
where $\delta n_e$ is the number density fluctuation of the free electrons in IGM, and $D$ is the distance between an FRB source and the observer. 
Note that even if $\delta n_e$ is taken as of the same order of $n_e$, i.e. $| \delta n_e| \sim n_e$, the DM variation is negligibly small.

\subsubsection{Density fluctuation in galaxy groups and clusters}
The LSS density fluctuation leads to the formations of galaxy clusters or groups. 
Galaxy clusters and groups are the largest known self-gravitationally bound systems in the process of cosmic structure formation, which form in the densest part of the LSS of the universe \citep[e.g.][]{bah99}. Galaxy groups are the small aggregates of galaxies. There are $3-30$ bright galaxies with a radius of $R\simeq(0.1-1)h^{-1}~\unit{Mpc}$. The typical random peculiar velocities of the galaxies in a group is $v_p\sim (100-500)~\unit{km~s^{-1}}$, which corresponds to a typical group mass of $\sim (10^{12.5}-10^{14})h^{-1} M_\odot$. The groups exhibit a spatial number density of $n_{\rm group}\sim(10^{-3}-10^{-5})h^3~\unit{Mpc^{-3}}$, and the fraction of galaxies in groups is $\sim55\%$ \citep[e.g.][]{bah99}. 
X-ray observations show that the central free electron density is about $n_{e,c}\lesssim10^{-3}~\unit{cm^{-3}}$ for the groups \citep[e.g.][]{kra12}.
Clusters are larger structures than groups, but there is no significant dividing line between clusters and groups. In general, there are $30-300$ bright galaxies with a radius of $R\simeq(1-2)h^{-1}~\unit{Mpc}$ in a cluster. The galaxies in a cluster has larger random peculiar velocities, e.g. $v_p\sim (400-1400)~\unit{km~s^{-1}}$, implying a typical cluster mass of $\sim (10^{14}-2\times10^{15})h^{-1} M_\odot$. The spatial number density of the clusters is $n_{\rm cluster}\sim(10^{-5}-10^{-6})h^3~\unit{Mpc^{-3}}$, and the galaxy fraction is only $\sim5\%$ \citep{bah99}. The central free electron density is about $n_{e,c}\sim10^{-3}~\unit{cm^{-3}}$ for clusters. 

In order to calculate the DM variation from the gas density fluctuation in groups or clusters,
we first consider the probability of an FRB passing through groups or clusters. We assume that the distance between the FRB source and the observer is $D$. The ``optical depth'' (or chance probability) is approximately 
\begin{eqnarray}
\tau_{\rm group}&\sim& n_{\rm group} (\pi R^2) D\simeq 0.03\left(\frac{n_{\rm group}}{10^{-3}h^3~\unit{Mpc^{-3}}}\right)\nonumber\\
&\times& \left(\frac{R}{0.1~h^{-1}~\unit{Mpc}}\right)^{2}\left(\frac{D}{1~\unit{Gpc}}\right) \nonumber\\
\end{eqnarray} 
to intersect a foreground group, and
\begin{eqnarray}
\tau_{\rm cluster}&\sim& n_{\rm cluster} (\pi R^2) D\simeq 0.03\left(\frac{n_{\rm cluster}}{10^{-5}h^3~\unit{Mpc^{-3}}}\right)\nonumber\\
&\times& \left(\frac{R}{1~h^{-1}~\unit{Mpc}}\right)^{2}\left(\frac{D}{1~\unit{Gpc}}\right)\nonumber\\
\end{eqnarray}
to intersect a foreground cluster. Both probabilities are small. However, it is possible that FRBs are born in groups or clusters, and any extragalactic light always pass through our Local Group.

The DM variation contributed by proper motion of galaxies in a galaxy group or cluster can be estimated as 
\begin{eqnarray}
\left| \frac{d\rm DM}{dt} \right|&\sim&\frac{|\delta n_e| v_{\rm flu} R}{l_{\rm flu}}\sim |\delta n_e| v_p\simeq5.1\times10^{-9}~\unit{pc~cm^{-3}yr^{-1}}\nonumber\\
&\times&\left(\frac{|\delta n_e|}{10^{-5}~\unit{cm^{-3}}}\right)\left(\frac{v_p}{500~\unit{km~s^{-1}}}\right),
\end{eqnarray}
where $\delta n_e$ is the density fluctuation of the electron gas, 
$l_{\rm flu}$ is the characteristic fluctuation scale in a cluster or group, $v_{\rm flu}$ is the characteristic fluctuation velocity of the hot gas in the cluster or group. The hydrodynamic simulations showed that $l_{\rm flu}\lesssim R$ and $v_{\rm flu}\sim v_p$ \citep[e.g.][]{gas14}.
Since the electron gas density profile in a group or cluster is approximately $n_e\propto r^{-2}$ for $0.1h^{-1}~\unit{Mpc^{-3}}<r<1.5h^{-1}~\unit{Mpc^{-3}}$, the mean electron number density is $n_e\sim0.03n_{e,c}$. Even if $\delta n_e$ is taken as of the same order of $n_e$, i.e. $| \delta n_e| \sim n_e$, the DM variation is negligibly small.

\subsection{Gas density fluctuation in the ISM}
The components of the interstellar medium include dense molecular gas, diffuse molecular gas, cold neutral medium, warm neutral medium, HII gas (including warm ionized medium and HII region), and coronal gas \citep[e.g.][]{dra11}. Among them, HII gas and coronal gas are ionized components. HII gas has a volume filling factor of $\sim 10\%$ and a gas density spanning a wide range, $(0.3-10^4)~\unit{cm^{-3}}$ \citep[e.g.][]{dra11}. An HII region is a dense cloud where hydrogen has been photoionized by ultraviolet photons from hot stars embedded in the cloud, 
and the extended low-density photoionized regions is referred to as the warm ionized medium. Coronal gas has a large fraction volume of $\sim50\%$ but an extremely low gas density of $\sim0.004~\unit{cm^{-3}}$ \citep[e.g.][]{dra11}. Therefore, the DM contribution in the ISM is mainly from the HII gas.

Similar to the discussion about the hot gas in clusters or groups, the DM variation from the ISM can be estimated as
\begin{eqnarray}
\left|\frac{d\rm DM_{ISM}}{dt} \right| \sim |\delta n_e v_p|&\simeq&2\times10^{-5}~\unit{pc~cm^{-3}yr^{-1}}\nonumber\\
&\times&\left(\frac{|\delta n_e|}{0.1~\unit{cm^{-3}}}\right)\left(\frac{|v_p|}{200~\unit{km~s^{-1}}}\right).\nonumber\\
\label{eq:DM-var-ISM}
\end{eqnarray}
where $v_p\sim200~\unit{km~s^{-1}}$ is taken as the typical rotation velocity of Milky Way. And even if $\delta n_e$ is taken as of the same order of the mean ionized gas density in ISM, e.g. $|\delta n_e|\sim0.1~\unit{cm^{-3}}$\citep[e.g.][]{dra11}, the DM variation is negligibly small. 
One can see that the DM variations due to a large-scale plasma, from the cosmological scale to the galaxy scale, cannot significantly change the  DM of an FRB during an observable time. 

In the following, we consider the DM variation caused by the plasma local to the FRB source, which includes a supernova remnant (SNR), a pulsar wind nebula (PWN), and a local HII region.

\subsection{SNR}

A supernova leaves behind an SNR, which has three phases:
(1) {\em the free-expansion phase}, during which the SNR ejecta moves with roughly the initial velocity, and the SNR mass is dominated by the ejecta; (2) {\em the Sedov-Taylor phase}, during which the ejecta begins to slow down due to the interaction with the ambient medium, and the SNR mass is dominated by the swept ambient medium; and (3) {\em the snowplow phase}, during which the shock wave undergoes significant radiative cooling. 

Note that the observed DM variation has two contributions: the first one is from the ionized medium (including ejecta and swept ambient medium) in SNR, e.g. $d{\rm DM_{SNR}}/dt$; the second one is due to the decrease of the ionized ambient medium in the upstream of the shock, e.g. $d{\rm DM_{ISM}}/dt$. Thus the observed DM satisfies
\begin{eqnarray}
\frac{d\rm DM_{obs}}{dt}=\frac{d\rm DM_{SNR}}{dt}+\frac{d\rm DM_{AM}}{dt},\label{vdm}
\end{eqnarray}
where
\begin{eqnarray}
\frac{d\rm DM_{AM}}{dt}=-\eta_0nv,
\end{eqnarray}
$n$ is the number density of the ambient medium, $v$ is the SNR velocity, and $\eta_0$ is the ionization fraction in the ambient medium. For fully ionized medium, $\eta_0=1$; for neutral medium, $\eta_0=0$. We assume that the hydrogen is the dominant component of the ambient medium.

At first, we consider the dispersion of electromagnetic wave in the SNR in the early stage. Since the number density of the ionized medium in the ejecta is very large at this moment, the frequency-dependent delay time, $\Delta t\propto \nu^{-\alpha}$, might deviates from $\Delta t\propto \nu^{-2}$. Once $\alpha$ is measured, the electron number density can be constrained via \citep[e.g.][]{den14,tun14,kat16b}
\begin{eqnarray}
n_e&<&\frac{2}{3}\left|\alpha-2\right|\frac{m_e\omega^2}{4\pi e^2}\simeq2.5\times10^7~\unit{cm^{-3}}\nonumber\\
&\times&\left(\frac{|\alpha-2|}{0.003}\right)\left(\frac{\nu}{1~\unit{GHz}}\right)^2.\label{nc}
\end{eqnarray}
The tightest observational upper bound on $|\alpha-2|$ is 0.003 \citep[e.g.][]{kat16b}. During the free-expansion phase, the mass contribution from swept ambient medium can be neglected. For the thin shell approximation, the number density of the electron in the shell is given by
\begin{eqnarray}
n_e&\simeq&\frac{\eta M}{4\pi\mu_{m} m_p R^2\Delta R}=\frac{\eta M^{5/2}}{8\sqrt{2}\pi\mu_{m} m_p E_0^{3/2}t^3}\left(\frac{\Delta R}{R}\right)^{-1}\nonumber\\
&\simeq&2.5\times10^7~\unit{cm^{-3}}\eta\left(\frac{M}{M_\odot}\right)^{5/2}\left(\frac{t}{1~\unit{yr}}\right)^{-3}\nonumber\\
&\times&\left(\frac{E_0}{10^{51}~\unit{erg}}\right)^{-3/2}\left(\frac{\Delta R}{0.1 R}\right)^{-1},\label{ne}
\end{eqnarray}
where $M$ is the SNR mass, $E_0$ is the kinetic energy of the SNR, $t$ is the age of the SNR, $\eta$ is the ionization fraction of the medium in the SNR, and $\mu_m=1.2$ is the mean molecular weight for a solar composition in the SNR ejecta. 
The free-expansion velocity of the SNR is estimated as $v\simeq(2E_0/M)^{1/2}\simeq10000~\unit{km~s^{-1}}(M/M_\odot)^{-1/2}(E_0/10^{51}~\unit{erg})^{1/2}$, and the SNR radius satisfies $R\sim v t\simeq0.01~\unit{pc}(M/M_\odot)^{-1/2}(E_0/10^{51}~\unit{erg})^{1/2}(t/1~\unit{yr})$. 
According to Eq.(\ref{nc}) and Eq.(\ref{ne}), one can constrain the lower limit of SNR age via the observed frequency-dependent delay time ($\Delta t\propto\nu^{-\alpha}$), e.g.
\begin{eqnarray}
t&>&1~\unit{yr}~\eta^{1/3}\left(\frac{M}{M_\odot}\right)^{5/6}\left(\frac{E_0}{10^{51}~\unit{erg}}\right)^{-1/2}\left(\frac{\Delta R}{0.1 R}\right)^{-1/3}\nonumber\\
&\times&\left(\frac{|\alpha-2|}{0.003}\right)^{-1/3}\left(\frac{\nu}{1~\unit{GHz}}\right)^{-2/3}.
\end{eqnarray}

The DM variation of an SNR in the free-expansion phase has been studied by various authors \citep[e.g.][]{kat16,pir16,met17,yan17,pir17,kat17}. A time dependent DM provided by a young SNR in the free-expansion phase can be estimated as
\begin{eqnarray}
{\rm DM_{SNR,FE}}&\simeq&n_e\Delta R=\frac{\eta M^2}{8\pi \mu_m m_p E_0 t^2}\simeq260~\unit{pc~cm^{-3}}\nonumber\\
&\times&\eta\left(\frac{M}{M_\odot}\right)^2\left(\frac{t}{10~\unit{yr}}\right)^{-2}\left(\frac{E_0}{10^{51}~\unit{erg}}\right)^{-1},\nonumber\\
\end{eqnarray}
And the DM variation of the SNR is given by
\begin{eqnarray}
\frac{d\rm DM_{SNR,FE}}{dt}&=&-\frac{\eta M^2}{4\pi \mu_m m_p t^3E_0}=-52~\unit{pc~cm^{-3}yr^{-1}}\nonumber\\
&\times&\eta\left(\frac{M}{M_\odot}\right)^2\left(\frac{t}{10~\unit{yr}}\right)^{-3}\left(\frac{E_0}{10^{51}~\unit{erg}}\right)^{-1}.\nonumber\\
\end{eqnarray}
On the other hand, due to $|d{\rm DM_{AM}}/dt|\ll |d{\rm DM_{SNR,FE}}/dt|$ during free-expansion phase, the observed DM variation is dominated by the SNR contribution.
One can see that a young SNR with its age $\lesssim 10~\unit{yr}$ can give a significant decreasing DM with time. 

As more and more ambient medium is swept into the shock, the shock starts to decelerate at a radius
\begin{eqnarray}
R_{\rm dec}&=&\left(\frac{3M}{4\pi m_pn}\right)^{1/3}\nonumber\\
&\simeq&0.46~\unit{pc}\left(\frac{M}{M_\odot}\right)^{1/3}\left(\frac{n}{100~\unit{cm^{-3}}}\right)^{-1/3}.
\end{eqnarray}
The corresponding decelerated time is given by
\begin{eqnarray}
t_{\rm dec}&\simeq&\frac{R_{\rm dec}}{v}\nonumber\\
&\simeq&45~\unit{yr}\left(\frac{M}{M_\odot}\right)^{5/6}\left(\frac{E_0}{10^{51}~\unit{erg}}\right)^{-1/2}\left(\frac{n}{100~\unit{cm^{-3}}}\right)^{-1/3}.\nonumber\\
\end{eqnarray}
When $t\sim t_{\rm dec}$, the free-expansion phase is over, and the SNR enters the Sedov-Taylor phase. 
The SNR radius during the Sedov-Taylor phase satisfies the self-similar solution, which is given by \citep[e.g.][]{tay50,sed59,dra11}
\begin{eqnarray}
R_{\rm ST}&=&1.15 E_0^{1/5}\rho^{-1/5}t^{2/5}\simeq0.84~\unit{pc}\nonumber\\
&\times&\left(\frac{E_0}{10^{51}~\unit{erg}}\right)^{1/5}\left(\frac{n}{100~\unit{cm^{-3}}}\right)^{-1/5}\left(\frac{t}{100~\unit{yr}}\right)^{2/5},\nonumber\\
\end{eqnarray}
for $t\gg t_{\rm dec}$. 
The corresponding SNR velocity during the Sedov-Taylor phase reads
\begin{eqnarray}
v_{\rm ST}&=&\frac{dR_{\rm ST}}{dt}\simeq3280~\unit{km~s^{-1}}\nonumber\\
&\times&\left(\frac{E_0}{10^{51}~\unit{erg}}\right)^{1/5}\left(\frac{n}{100~\unit{cm^{-3}}}\right)^{-1/5}\left(\frac{t}{100~\unit{yr}}\right)^{-3/5},\nonumber\\
\end{eqnarray}
According to the strong shock jump condition, the temperature in the downstream medium of the shock is given by \citep[e.g.][]{dra11}
\begin{eqnarray}
T_{\rm ST}&\simeq&\frac{2(\gamma-1)}{(\gamma+1)^2}\frac{m_pv_{\rm ST}^2}{k}=\frac{3}{16}\frac{m_pv_{\rm ST}^2}{k}\simeq2.4\times10^8~\unit{K}\nonumber\\
&\times&\left(\frac{E_0}{10^{51}~\unit{erg}}\right)^{2/5}\left(\frac{n}{100~\unit{cm^{-3}}}\right)^{-2/5}\left(\frac{t}{100~\unit{yr}}\right)^{-6/5},\nonumber\\\label{tem}
\end{eqnarray}
for the adiabatic index $\gamma=5/3$. Due to $T_{\rm ST}\gg T_{\rm H}\sim1.6\times10^5~\unit{K}$, where $T_{\rm H}$ is the ionization temperature of hydrogen, all medium in the downstream of the shock would be ionized. The corresponding DM can be estimated as
\begin{eqnarray}
{\rm DM_{SNR,ST}}&\simeq&\int_0^{R_{\rm ST}} n(r)dr=\int_0^1\frac{\gamma+1}{\gamma-1}nf(x)R_{\rm ST}dx\nonumber\\
&=&\zeta n R_{\rm ST}\simeq34~\unit{pc~cm^{-3}}\left(\frac{E_0}{10^{51}~\unit{erg}}\right)^{1/5}\nonumber\\
&\times&\left(\frac{n}{100~\unit{cm^{-3}}}\right)^{4/5}\left(\frac{t}{100~\unit{yr}}\right)^{2/5},\nonumber\\
\end{eqnarray}
where $x=r/R_{\rm ST}$ is the dimensionless radius, $n(r)$ is the density profile in the shock downstream, $\zeta\equiv[(\gamma+1)/(\gamma-1)]\int f(x)dx \simeq0.4$ for the adiabatic index $\gamma=5/3$, $f(x)$ is the density dimensionless function of the Sedov-Taylor solution \citep[e.g.][]{dra11,tay50,sed59}. The DM variation of the SNR is given by
\begin{eqnarray}
\frac{d\rm DM_{SNR,ST}}{dt}&\simeq&+0.14~\unit{pc~cm^{-3}yr^{-1}}\nonumber\\
&\times&\left(\frac{E_0}{10^{51}~\unit{erg}}\right)^{1/5}\left(\frac{n}{100~\unit{cm^{-3}}}\right)^{4/5}\left(\frac{t}{100~\unit{yr}}\right)^{-3/5}.\nonumber\\
\end{eqnarray}
On the other hand, with the expansion of SNR, ambient medium outside $R_{\rm ST}$ becomes less. The DM variation contributed by the ambient medium is given by
\begin{eqnarray}
\frac{d\rm DM_{AM}}{dt}&=&-\eta_0nv_{\rm ST}=-\frac{\eta_0}{\zeta}\frac{d\rm DM_{SNR,ST}}{dt}.
\end{eqnarray}
Therefore, according to Eq.(\ref{vdm}), the observed DM variation is given by
\begin{eqnarray}
\frac{d\rm DM_{obs}}{dt}&=&0.35(0.4-\eta_0)~\unit{pc~cm^{-3}yr^{-1}}\nonumber\\
&\times&\left(\frac{E_0}{10^{51}~\unit{erg}}\right)^{1/5}\left(\frac{n}{100~\unit{cm^{-3}}}\right)^{4/5}\left(\frac{t}{100~\unit{yr}}\right)^{-3/5}.\nonumber\\
\end{eqnarray} 

Interestingly, different from the free-expansion phase, the DM evolution during the Sedov-Taylor phase depends on the ionization fraction $\eta_0$ of the ambient medium. If $\eta_0<0.4$, the DM will increase with time. The reason is as follows: as more and more ambient medium is swept into the shock, they will be ionized in the downstream, leading to the column density of free electrons in the SNR increasing. However, the DM variation contributed by the ambient medium in the upstream is small, because the ambient medium is near-neutral. 
On the other hand, if $\eta_0>0.4$, the DM will decrease with time, since the density profile of the all ionized ambient medium (including swept and non-swept medium) significantly changes after the SNR shock sweeping \footnote{For example, we assume that the ambient medium with number density $n$ are fully ionized, e.g. $\eta_0=1$, and the ejecta mass could be neglect. Before the supernova explosion, the DM within a radius $R$ satisfies ${\rm DM}\sim nR$. After the SNR reaching $R$, one has ${\rm DM}\sim(4\pi nR^3/3)\Delta R/(4\pi R^2\Delta R)\sim(1/3)nR$, where $\Delta R$ is the thickness of the SNR. Therefore, the DM changes $\Delta {\rm DM}\sim-(2/3)nR$. Here, we assume that $\Delta R\ll R$. The more precise conclusion is based on Sedov-Taylor solution, as discussed above.}.
Finally, we note that the DM variation could be observable during a long observational time, if the ambient medium is dense with $n\gtrsim100~\unit{cm^{-3}}$. 

Due to the radiative cooling, the SNR will leave the Sedov-Taylor phase and enter the snowplow phase, when the radiative energy is of the same order as the SNR kinetic energy. The radiative cooling time is approximately given by \citep[e.g.][]{dra11}
\begin{eqnarray}
t_{\rm rad}\simeq3920~\unit{yr}\left(\frac{E_0}{10^{51}~\unit{erg}}\right)^{0.22}\left(\frac{n}{100~\unit{cm^{-3}}}\right)^{-0.55}.
\end{eqnarray}
The SNR radius during the snowplow phase is given by \citep[e.g.][]{dra11}
\begin{eqnarray}
R_{\rm SP}&\simeq&R_{\rm ST}(t_{\rm rad})\left(\frac{t}{t_{\rm rad}}\right)^{2/7}\nonumber\\
&\simeq&3.6~\unit{pc}\left(\frac{t}{t_{\rm rad}}\right)^{2/7}\left(\frac{E_0}{10^{51}~\unit{erg}}\right)^{0.288}\left(\frac{n}{100~\unit{cm^{-3}}}\right)^{-0.42}.\nonumber\\
\end{eqnarray}
The corresponding SNR velocity reads
\begin{eqnarray}
v_{\rm SP}&=&\frac{dR_{\rm SP}}{dt}=260~\unit{km~s^{-1}}\nonumber\\
&\times&\left(\frac{t}{t_{\rm rad}}\right)^{-5/7}\left(\frac{E_0}{10^{51}~\unit{erg}}\right)^{0.288}\left(\frac{n}{100~\unit{cm^{-3}}}\right)^{-0.42}.\nonumber\\
\end{eqnarray}
Similar to Eq.(\ref{tem}), one has the temperature in the downstream medium during the snowplow phase, e.g.
\begin{eqnarray}
T_{\rm SP}&\simeq&\frac{3}{16}\frac{m_pv_{\rm SP}^2}{k}=1.5\times10^6~\unit{K}\nonumber\\
&\times&\left(\frac{t}{t_{\rm rad}}\right)^{-10/7}\left(\frac{E_0}{10^{51}~\unit{erg}}\right)^{0.576}\left(\frac{n}{100~\unit{cm^{-3}}}\right)^{-0.84}.\nonumber\\
\end{eqnarray}
Note that $T> T_{\rm H}$ for $t_{\rm rad}<t<5t_{\rm rad}$ and $T< T_{\rm H}$ for $t>5t_{\rm rad}$. Duing the snowplow phase, the ionization fraction $\eta$ will depend on the temperature in the SNR and the injection rate of ionizing photon from other stars.
Since the mass in the SNR is dominated by the swept ambient medium, the DM of SNR may be given by
\begin{eqnarray}
{\rm DM_{SNR,SP}}&\simeq&\frac{\eta M_{\rm sw}}{4\pi m_p R_{\rm SP}^2}\simeq\frac{1}{3}\eta nR_{\rm SP}=120~\unit{pc~cm^{-3}}\nonumber\\
&\times&\eta \left(\frac{t}{t_{\rm rad}}\right)^{2/7}\left(\frac{E_0}{10^{51}~\unit{erg}}\right)^{0.288}\left(\frac{n}{100~\unit{cm^{-3}}}\right)^{0.58},\nonumber\\
\end{eqnarray}
where $M_{\rm sw}=(4\pi/3)nm_pR_{\rm SP}^3$ is the mass of the swept ambient medium, which is much larger than the original mass of the ejecta. The DM variation is given by
\begin{eqnarray}
\frac{d\rm DM_{SNR,SP}}{dt}&\simeq&+0.009~\unit{pc~cm^{-3}yr^{-1}}\nonumber\\
&\times&\left(\frac{t}{t_{\rm rad}}\right)^{-5/7}\left(\frac{E_0}{10^{51}~\unit{erg}}\right)^{0.068}\left(\frac{n}{100~\unit{cm^{-3}}}\right)^{1.13}.\nonumber\\
\end{eqnarray}
On the other hand, the DM variation contributed by the ambient medium is given by
\begin{eqnarray}
\frac{d\rm DM_{AM}}{dt}&=&-\eta_0nv_{\rm SP}=-\frac{3\eta_0}{\eta}\frac{d\rm DM_{SNR,SP}}{dt}
\end{eqnarray}
Therefore, the observed DM variation is given by
\begin{eqnarray}
\frac{d\rm DM_{obs}}{dt}&\simeq&0.009~\unit{pc~cm^{-3}yr^{-1}}(1-\frac{3\eta_0}{\eta})\nonumber\\
&\times&\left(\frac{t}{t_{\rm rad}}\right)^{-5/7}\left(\frac{E_0}{10^{51}~\unit{erg}}\right)^{0.068}\left(\frac{n}{100~\unit{cm^{-3}}}\right)^{1.13}.\nonumber\\
\end{eqnarray}
Similar to the Sedov-Taylor phase, the DM evolution during the snowplow phase depends on the relation between $\eta_0$ and $\eta$ of the ambient medium. If $\eta_0<\eta/3$, the DM will increase with time, otherwise, the DM will decrease with time.

\subsection{PWN}

Many FRB models invoke a young pulsar or magnetar as the source of the bursts \citep{pop10,kul14,con16,cor16,yan16b,mur16,pir16,bel17,mur17,met17,cao17,dai17,nic17,wax17}. The pulsar/magnetar parameters have been constrained with the available observations for FRB 121102 \citep{ten16,lyu17,zha17}, but generally the data require a relatively young pulsar, which is supposed to have a surrounding PWN.
We now consider the DM contribution from a PWN. In a PWN, the local magnetic field in the wind may not be neglected, since the electron cyclotron frequency is larger than the wave frequency. One should consider wave dispersion in a magnetized plasma.
In a magnetized plasma, there are four modes of electromagnetic waves \citep[e.g.][]{mes92}, i.e. O mode ($\overrightarrow{k}\perp\overrightarrow{B},\overrightarrow{E}_{\rm w}\parallel\overrightarrow{B}$), X mode ($\overrightarrow{k}\perp\overrightarrow{B},\overrightarrow{E}_{\rm w}\perp\overrightarrow{B}$), R mode ($\overrightarrow{k}\parallel\overrightarrow{B}$, right circular polarization) and L mode ($\overrightarrow{k}\parallel\overrightarrow{B}$, left circular polarization), 
where $\overrightarrow{k}$ is the wave vector of electromagnetic wave, $\overrightarrow{B}$ is the local magnetic field, $\overrightarrow{E}_{\rm w}$ is the electric field vector of the electromagnetic wave. Their dispersion relations read \citep[e.g.][]{mes92}
\begin{eqnarray}
1-\frac{\omega_p^2}{\omega^2}=\frac{c^2k^2}{\omega^2},~~~~~&{\rm for~O~mode;}&\label{dr}\\
1-\frac{\omega_p^2}{\omega^2}\frac{\omega^2-\omega_p^2}{\omega^2-\omega_h^2}=\frac{c^2k^2}{\omega^2},~~~~~&{\rm for~X~mode;}&\label{drx}\\
1-\frac{\omega_p^2/\omega^2}{1-\omega_c/\omega}=\frac{c^2k^2}{\omega^2},~~~~~&{\rm for~R~mode;}&\\
1-\frac{\omega_p^2/\omega^2}{1+\omega_c/\omega}=\frac{c^2k^2}{\omega^2},~~~~~&{\rm for~L~mode;}&
\end{eqnarray}
respectively, where $\omega_c=eB/m_ec$ is the electron cyclotron frequency, and $\omega_h=(\omega_p^2+\omega_c^2)^{1/2}$ is the upper hybrid frequency. Here, we neglect the motion of ions due to their large masses. 

In the magnetized plasma, the two important properties are the presence of resonances and cutoffs, which occur at $k\rightarrow\infty$ and $k\rightarrow 0$, respectively.
In O mode, the wave dispersion relation is the same as that in an unmagnetized plasma and its cutoff occurs at $\omega=\omega_p$. Only the O wave with $\omega>\omega_p$ can propagate through the medium. 
In X mode, the wave has a more complicated dispersion relation. The resonance occurs at $\omega=\omega_h$ and the cutoffs occur at $\omega=\omega_L$ and $\omega=\omega_R$, where
$\omega_{R,L}=\left[\pm\omega_c+(\omega_c^2+4\omega_p^2)^{1/2}\right]/2$. Only the X wave with $\omega_L<\omega<\omega_h$ or $\omega>\omega_R$ can propagate through the medium. 
In R mode, the wave is right circularly polarized, and its cutoff occurs at $\omega=\omega_R$ and resonance occurs at $\omega=\omega_c$. Only the R wave with $\omega>\omega_R$ or $0<\omega<\omega_c$ can propagate through the medium. 
In L mode, the wave is left circularly polarized, and the cutoff occurs at $\omega=\omega_L$ and it has no resonance. Only the L wave with $\omega>\omega_L$ can propagate through the medium. 
The dispersion relations of R mode and L mode approach the dispersion relation in an unmagnetized plasma only when $\omega\gg\omega_c$.

The radiation mechanism of the coherent radio emission from pulsars is still not identified. Observations of pulsars show that there might be at least two locations of radio emission, one in the open field line region near the polar cap, and the other near the light cylinder \citep[e.g.][]{mel17}. It is not clear whether any of those mechanisms might produce FRBs with even much higher brightness temperatures than pulsar radio emission. In any case, it is safe to integrate the DM contribution of a PWN starting from the light cylinder. In the pulsar magnetosphere, the dispersion relation of electromagnetic wave is very complex \citep[e.g.][]{aro86}. More importantly, the radio emission is likely still in the growth mode within the magnetosphere. This is certainly the case for the models invoking coherent emission near the light cylinder or the outer gap, but is valid for some polar cap models as well \citep[e.g.][]{mel17}. In the following, we only consider the DM contribution outside the light cylinder, which corresponds to a lower limit but should be close to the total DM near a pulsar.

For a magnetized rapidly-rotating neutron star, a relativistic electron-positron pair wind is expected to stream out from the magnetosphere \citep{mur16,cao17,dai17}. 
The pair number density of the wind at $r$ is given by \citep[e.g.][]{yu14}
\begin{eqnarray}
n_{\rm w}(r)\simeq\frac{\dot N_{\rm w}}{4\pi r^2c}=\mu_\pm n_{\rm GJ}(R_{\rm LC})\left(\frac{r}{R_{\rm LC}}\right)^{-2},
\end{eqnarray}
where $\mu_\pm$ is the multiplicity parameter of the electron-positron pairs, $R_{\rm LC}\equiv c/\Omega$ is the radius of the light cylinder, $n_{\rm GJ}=(\Omega B_p/2\pi ec)(r/R)^{-3}$ is the Goldreich-Julian number density, $\dot N_{\rm w}\simeq 4\pi R_{\rm LC}^2\mu_\pm n_{\rm GJ}(R_{\rm LC})c$ is the particle number flux, $R$ is the radius of the neutron star, and $B_p$ is magnetic field strength at the polar cap. In the PWN, the plasma frequency is given by
\begin{eqnarray}
\omega_p&=&\left(\frac{4\pi e^2n_{\rm w}}{m_e}\right)^{1/2}\simeq4.5\times10^9~\unit{rad~s^{-1}}\nonumber\\
&\times&\left(\frac{\mu_\pm}{10^4}\right)^{1/2}\left(\frac{B_p}{10^{14}~\unit{G}}\right)^{1/2}\left(\frac{P}{0.1~\unit{s}}\right)^{-2}\left(\frac{r}{R_{\rm LC}}\right)^{-1},
\end{eqnarray}
and the electron cyclotron frequency is given by
\begin{eqnarray} 
\omega_c&\simeq&\frac{eB_p}{m_ec}\left(\frac{R_{\rm LC}}{R}\right)^{-3}\left(\frac{r}{R_{\rm LC}}\right)^{-1}\simeq1.6\times10^{13}~\unit{rad~s^{-1}}\nonumber\\
&\times&\left(\frac{B_p}{10^{14}~\unit{G}}\right)\left(\frac{P}{0.1~\unit{s}}\right)^{-3}\left(\frac{r}{R_{\rm LC}}\right)^{-1}.
\end{eqnarray}
Note that at $r>R_{\rm LC}$, $B\propto r^{-1}$, so that the factor of $(\omega_p/\omega_c)^2$ is independent of radius. Near the magnetosphere, the composition of wave modes and the conversion among them are complex, which depend on the radiation mechanism, magnetic field configuration, and the plasma environment in strong magnetic fields. 
For simplification, we assume that the angles between the wave polarization and $\overrightarrow{k}$ and $\overrightarrow{B}$ are essentially unchanged during the wave propagation, and ignore any conversion among modes. Considering more complicated situations does not alter the basic picture discussed below.

First, O-mode wave satisfies the classical unmagnetized-plasma dispersion relation, so that 
the corresponding DM contributed by the pulsar/magnetar wind may be given by \citep{yu14,cao17,yan17}
\begin{eqnarray}
{\rm DM_{w,O}}&\simeq&\int_{R_{\rm LC}}^{R_{\rm sh}} 2\Gamma(r)n_{\rm w}(r)dr\simeq3\Gamma_{\rm LC}\mu_\pm n_{\rm GJ}(R_{\rm LC})R_{\rm LC}\nonumber\\
&\simeq&380~\unit{pc~cm^{-3}}\left(\frac{\mu_\pm}{10^4}\right)^{2/3}\left(\frac{B_p}{10^{14}~\unit{G}}\right)^{4/3}\left(\frac{P}{0.1~\unit{s}}\right)^{-11/3}\nonumber\\
\end{eqnarray}
for $R_{\rm sh}\gg R_{\rm LC}$, where $P$ is the rotation period of the neutron star, $R_{\rm sh}$ is the radius of the shock (where the pulsar wind collides with the SNR \citep{mur16,mur17,kas17} or an ambient interstellar medium \citep{dai17}), $\Gamma_{\rm LC}\sim (L_{\rm sd}/\dot N_{\rm w}m_ec^2)^{1/3}$ is Lorentz factor of the relativistic wind at $R_{\rm LC}$, and $L_{\rm sd}=B_p^2R^6\Omega^4/6c^3$ is the spin-down luminosity of the pulsar/magnetar. The wind Lorentz factor at radius $r$ may be given by $\Gamma(r)\sim\Gamma_{\rm LC}L(r/R_{\rm LC})^{1/3}$ \citep{dre02}. 
In fact, the contribution of ${\rm DM_{w,O}}$ is mainly from the inner region, due to the larger number density of the electrons.

For X-mode wave,
near the light cylinder $\omega\ll\omega_p\ll\omega_c$ is satisfied. According to Eq.(\ref{drx}), the dispersion relation reads $c^2k^2/\omega^2\simeq1-\omega_{p,{\rm eff}}^2/\omega^2$, where $\omega_{p,{\rm eff}}=\omega_p^2/\omega_c\sim\omega_L\ll\omega_p$ is the effective plasma frequency. In this case, the dispersion relation of the X wave is similar to that in an unmagnetized plasma (with $\omega_{p,{\rm eff}}$ replaced by $\omega_p$), and the condition for the propagation of the X-mode wave is $\omega>\omega_{p,{\rm eff}}$. Therefore, the apparent DM is corrected by a factor $(\omega_p/\omega_c)^2$, i.e.
\begin{eqnarray}
{\rm DM_{w,X}}&\simeq&\left(\frac{\omega_p}{\omega_c}\right)^2{\rm DM_{w,O}}\simeq3\times10^{-5}~\unit{pc~cm^{-3}}\nonumber\\
&\times&\left(\frac{\mu_\pm}{10^4}\right)^{5/3}\left(\frac{B_p}{10^{14}~\unit{G}}\right)^{1/3}\left(\frac{P}{0.1~\unit{s}}\right)^{-5/3}.\nonumber\\
\end{eqnarray}
Therefore, the GHz X-mode wave can still propagate through the medium even though $\omega<\omega_p$ \citep[e.g.][]{kum17}, and the apparent DM contribution is extremely small. However, as the X wave propagates further, e.g. $r\gtrsim r_p$, where $r_p$ is defined by $\omega_p(r)=\omega$, the dispersion relation of X wave would not keep the unmagnetized form (see Eq.(\ref{drx})). In particular, when $\omega_p\ll\omega\ll\omega_c$, the dispersion relation is approximately $c^2k^2/\omega^2\simeq1+\omega_p^2/\omega_c^2$, leading to the group velocity satisfying $v_g=\partial\omega/\partial k\simeq c/(1+\omega_p^2/\omega_c^2)^{1/2}$. Thus, the electromagnetic wave with $\omega_p\ll\omega\ll\omega_c$ does not show significantly frequency-dependent delay time, i.e. there is no contribution to DM.  Finally, as the X wave propagates further, the resonance occurs when $\omega=\omega_h\sim\omega_c$ is satisfied. 

For R-mode and L-mode wave, near the light cylinder $\omega\ll\omega_c$ is satisfied. Their dispersion relations could be approximated as $c^2k^2/\omega^2=1\pm\omega_p^2/(\omega_c\omega)$, where ``$+$'' and ``$-$'' correspond to R mode and L mode, respectively. The group velocity is $v_g=c(1\pm\omega_{p,{\rm eff}}/\omega)^{1/2}/(1\pm\omega_{p,{\rm eff}}/2\omega)$, where $\omega_{p,{\rm eff}}=\omega_p^2/\omega_c$. Due to $\omega\gg\omega_{p,{\rm eff}}$, the group velocities of both modes are extremely close to the speed of light, i.e., $v_g\simeq c$. There is no significant frequency-dependent delay time. As the waves propagate further, both $\omega_{p}$ and $\omega_c$  decrease. For R-mode wave, the resonance finally occurs at $\omega=\omega_c$; For L-mode wave, its dispersion relation approaches that in the unmagnetized plasma when $\omega\gg\omega_c$, but the DM contribution from the PWN is very small at that time.

Besides the wind contribution, the shocked electron-positron pairs may contribute to a part of the dispersion measure. These thermalized electron-positron pairs would undergo cooling and become non-relativistic (if the radiative cooling time is much less than that of the dynamical time of the shock). In this case, due to $\omega\gg\omega_c$, the dispersion relations of all modes keep the classical unmagnetized form. 
The dispersion measure from these thermalized particles is given by \citep{yan17}
\begin{eqnarray}
{\rm DM_{sh}}&\simeq&\frac{\dot N_{\rm w}T_{\rm sd}}{4\pi R_{\rm sh}^2}\simeq
\frac{3c^2\mu_\pm I}{2\pi eB_pR^3R_{\rm sh}^2}
\simeq3\times10^{-7}\unit{pc~cm^{-3}}\nonumber\\
&\times&\left(\frac{\mu_\pm}{10^4}\right)\left(\frac{B_p}{10^{14}~\unit{G}}\right)^{-1}\left(\frac{R_{\rm sh}}{0.1~\unit{pc}}\right)^{-2},
\end{eqnarray}
for $T\gg T_{\rm sd}$, where $T$ is the PWN age, $T_{\rm sd}$ is the spindown time of the pulsar, $I\simeq 10^{45}~\unit{g~cm^2}$ is the moment of inertia, $\dot N_{\rm w}T_{\rm sd}$ is the total number of electron-positron pairs ejected over $T_{\rm sd}$, which does not depend on $\Omega$.
Due to ${\rm DM_{sh}}\ll{\rm DM_{w,O}}$, one has ${\rm DM_{PWN}}\simeq{\rm DM_{w,O}}$ \citep{yan17}.
Therefore, the DM variation of the PWN wind may be given by
\begin{eqnarray}
\frac{{d\rm DM}_{\rm w}}{dt}&\simeq&\frac{d{\rm DM_{w,O}}}{dP}\frac{dP}{dt}
\simeq-11~\unit{pc~cm^{-3}yr^{-1}}\nonumber\\
&\times&\left(\frac{\mu_\pm}{10^4}\right)^{2/3}\left(\frac{B_p}{10^{14}~\unit{G}}\right)^{10/3}\left(\frac{P}{0.1~\unit{s}}\right)^{-17/3},
\end{eqnarray}
where $dP/dt=(2\pi^2/3c^3)(B^2R^6/IP)$. Here, we only consider the contribution from the O-mode waves, since the DMs of the other modes are extremely small. One can see that a young pulsar/magnetar wind may provide a significant contribution to DM and to its variation.

\subsection{HII region}

Next, we consider an FRB embedded in an HII region in its host galaxy. If the FRB source is related to recent star formation (as indicated by the host galaxy type of FRB 121102), there are two (exotic) mechanisms to cause DM variation. One is that there might be a new-born young star in the HII region, whose ionization front still evolves with time. The other possibility is that the FRB source may be rapidly moving within the HII region due to, e.g. a supernova kick.  In general, the evolution of the HII region radius $R$ may be estimated as
\begin{eqnarray}
N_u=4\pi nR^2\frac{dR}{dt}+\alpha_{\rm B}n^2\frac{4\pi}{3}R^3,\label{h2}
\end{eqnarray}
where $N_u$ is the emission rate of the ionizing photons from a star, $n$ is the gas number density in the HII region, and $\alpha_{\rm B}$ is the recombination rate. The first term represents ionization of the surrounding neutral gas that absorbs ionizing photon. The second term represents recombination of the ionized gas. Here we have assumed that the neutral gas is the neutral atomic hydrogen.
Eq.(\ref{h2}) has the solution 
\begin{eqnarray}
R(t)=R_{\rm str}\left(1-e^{-\alpha_{\rm B}nt}\right)^{1/3}.\label{rt}
\end{eqnarray}
Once $t\gg 1/(\alpha_Bn)$, the radius of the HII region reaches the maximum value $R\simeq R_{\rm str}\equiv\left(3N_u/4\pi\alpha_{\rm B} n^2\right)^{1/3}$,
where $R_{\rm str}$ is defined as Str\"omgren radius \citep{str39}. In this case, the ionization and recombination are balanced. 

We consider the case that FRB is associated with an HII region that is initially expanding outward due to the ionizing photon injection. 
For simplification, we assume that the timescale for the onset of a star is short. In this case, the star has just started producing a large flux of ionizing photons, making an expanding HII region.
According to Eq.(\ref{h2}) or Eq.(\ref{rt}), for $t\ll 1/(\alpha_{\rm B} n)\simeq1220~\unit{yr}(n/100~\unit{cm^{-3}})^{-1}$, where $\alpha_{\rm B}=2.6\times10^{-13}~\unit{cm^3s^{-1}}$ for $T=10^{4}~\unit{K}$, recombination can be neglected. The radius of the HII region is then approximately given by
\begin{eqnarray}
R(t)&=&\left(\frac{3N_u}{4\pi n}t\right)^{1/3}
\simeq2.3~\unit{pc}\left(\frac{N_u}{5\times10^{49}~\unit{s}^{-1}}\right)^{1/3}\nonumber\\
&\times&\left(\frac{n}{100~\unit{cm^{-3}}}\right)^{-1/3}\left(\frac{t}{100~\unit{yr}}\right)^{1/3}.
\end{eqnarray}
Here we assume that there is an O5 star with $N_u\sim5\times10^{49}~\unit{s^{-1}}$ in the HII region.
Therefore, the time-dependent DM of a young ($t\ll1/(\alpha_{\rm B} n)$) HII region is given by
\begin{eqnarray}
{\rm DM_{HII}}&=&nR(t)\simeq230~\unit{pc~cm^{-3}}\left(\frac{N_u}{5\times10^{49}~\unit{s}^{-1}}\right)^{1/3}\nonumber\\
&\times&\left(\frac{n}{100~\unit{cm^{-3}}}\right)^{2/3}\left(\frac{t}{100~\unit{yr}}\right)^{1/3},
\end{eqnarray}
and the corresponding DM variation is given by
\begin{eqnarray}
\frac{d\rm DM_{HII}}{dt}&=&\left(\frac{N_un^2}{36\pi t^2}\right)^{1/3}
\simeq+0.78~\unit{pc~cm^{-3}yr^{-1}}\nonumber\\
&\times&\left(\frac{N_u}{5\times10^{49}~\unit{s}^{-1}}\right)^{1/3}\left(\frac{n}{100~\unit{cm^{-3}}}\right)^{2/3}\left(\frac{t}{100~\unit{yr}}\right)^{-2/3}.\nonumber\\
\end{eqnarray} 
One can see that a young HII region can provide an observable positive DM variation.

For an old HII region with $t\gg 1/(\alpha_{\rm B} n)$, due to the balance of the ionization and recombination, the HII region radius would keep at the Str\"omgren radius, i.e., $R\sim R_{\rm str}$ and $dR/dt\sim 0$. The DM contributed by a Str\"omgren sphere is time-independent, i.e. $d{\rm DM_{HII}}/dt\sim0$, with a value \citep{yan17}
\begin{eqnarray}
{\rm DM_{HII}}&\simeq&nR_{\rm str}=\left(\frac{3N_un}{4\pi\alpha_{\rm B}}\right)^{1/3}\nonumber\\
&=&540~\unit{pc~cm^{-3}}\left(\frac{N_u}{5\times10^{49}~\unit{s}^{-1}}\right)^{1/3}\left(\frac{n}{100~\unit{cm^{-3}}}\right)^{1/3}.\nonumber\\
\end{eqnarray}
For an O5 star in HII region, one has $\alpha_{\rm B}=2.6\times10^{-13}~\unit{cm^3s^{-1}}$ for $T=10^{4}~\unit{K}$, and $R_{\rm str}=5.4~\unit{pc}$. The scale of the ionized region is larger than the projected size of $\lesssim0.7~\unit{pc}$ of the persistent source of FRB 121102 \citep{mar17}. As discussed above, the balanced HII region cannot provide DM variation for a static source in the HII region. 

On the other hand, if the FRB source is moving in an HII region, the plasma in the HII region may give a change to a DM variation. There are two scenarios here: 1. An FRB source is inside the HII region and is moving toward or away from us, which would decrease or increase the line-of-sight DM contribution; 2. An FRB source moves in the transverse direction, so that it enters or exits an HII region and may cause a sudden change in the measured DM along the line of sight. For the first case, the DM variation is given by
\begin{eqnarray} 
\left| \frac{d\rm DM_{HII}}{dt} \right| &\sim& n | v_\parallel | \simeq0.02~\unit{pc~cm^{-3}yr^{-1}}\nonumber\\
&\times&\left(\frac{n}{100~\unit{cm^{-3}}}\right)\left(\frac{|v_\parallel |}{200~\unit{km~s^{-1}}}\right),
\end{eqnarray}
where $v_\parallel$ is the velocity along the line of sight. Note that the typical pulsar kick velocities are $v\lesssim300~\unit{km~s^{-1}}$ \citep{hob05}, and $v_\parallel<v$. For the second case, the DM variation may be given by
\begin{eqnarray}
\left| \frac{d\rm DM_{HII}}{dt} \right| &\sim& \frac{nR}{\Delta R/|v_\perp|} \simeq0.02~\unit{pc~cm^{-3}yr^{-1}}\nonumber\\
&\times&\left(\frac{n}{100~\unit{cm^{-3}}}\right)\left(\frac{|v_\perp |}{200~\unit{km~s^{-1}}}\right)\left(\frac{\Delta R}{R}\right)^{-1},\nonumber\\
\end{eqnarray}
where $v_\perp$ is the transverse velocity, and $\Delta R$ corresponds to the thickness of the HII region edge. Similar to the first case, $v_\perp<v\lesssim300~\unit{km~s^{-1}}$ \citep{hob05}.

It is worth commenting that the free-free absorption optical depth in the HII region is small, so that FRBs can escape. The free-free absorption coefficient in the Rayleigh-Jeans limit is given by
\begin{eqnarray}
\alpha_{\rm ff}&=&\frac{4}{3}\left(\frac{2\pi}{3}\right)^{1/2}\frac{Z^2e^6n_en_i\bar g_{\rm ff}}{c m_e^{3/2}(k_BT)^{3/2}\nu^2},\nonumber\\
\bar g_{\rm ff}&=&\frac{\sqrt 3}{\pi}\left[\ln\left(\frac{(2k_BT)^{3/2}}{\pi e^2m_e^{1/2}\nu}\right)-\frac{5}{2}\gamma\right],
\end{eqnarray}
where $\gamma=0.577$ is Euler's constant, $\bar g_{\rm ff}$ is the Gaunt factor, and $n_i$ and $n_e$ are the number densities of ions and electrons, respectively. Here we take $n_e=n_i$ and $Z=1$ for an HII region. The optical depth for the free-free absorption is given by \citep[e.g.][]{lua14,mur16,met17,yan17}
\begin{eqnarray}
\tau_{\rm ff}\lesssim\alpha_{\rm ff} R_{\rm str}&\simeq&0.018\left(\frac{n}{100~\unit{cm^{-3}}}\right)^{4/3}\nonumber\\
&\times&\left(\frac{T}{10^4~\unit{K}}\right)^{-1.5}\left(\frac{\nu}{1~\unit{GHz}}\right)^{-2},
\end{eqnarray}
where $\alpha_{\rm B}=2.6\times10^{-13}~\unit{cm^3~s^{-1}}$ and $\bar g_{\rm ff}=6.0$ for $T=10^4~\unit{K}$ and $\nu=1~\unit{GHz}$ are taken, and the HII region radius satisfies $R\lesssim R_{\rm str}$. Therefore, such an HII region is optically thin for FRBs.

\subsection{Plasma lensing}

As pointed out by \citet{cor17}, plasma lenses in FRB host galaxies can generate multiple images for one burst, which
may arrive differentially by $<1\unit{\mu s}$ to tens of $\unit{ms}$ and show an apparent DM perturbation of a few $\unit{pc~cm^{-3}}$.
Here, we briefly discuss the perturbation and evolution of DM contributed by a plasma lens. We assume that the source-lens, source-observer, and lens-observer distances are $d_{\rm sl}$, $d_{\rm so}$, and $d_{\rm lo}=d_{\rm so}-d_{\rm sl}$, respectively. The transverse coordinates in the source, lens, and observer's planes are $x_{\rm s}$, $x$ and $x_{\rm o}$, respectively. 
We define $u_{\rm s}=x_{\rm s}/a$, $u=x/a$ and $u_{\rm o}=x_{\rm o}/a$, respectively, and define the effective offset as
\begin{eqnarray}
u^\prime\equiv\frac{d_{\rm lo}}{d_{\rm so}}u_{\rm s}+\frac{d_{\rm sl}}{d_{\rm so}}u_{\rm o}.\label{up}
\end{eqnarray}
Assuming that the lens centered on $u=0$ has a characteristic scale $a$, and the DM profile of the lens has the form ${\rm DM}(u)={\rm DM_l}\exp(-u^2)$, one can derive the lens equation in geometric optics \citep{cle98,cor17}, e.g.
\begin{eqnarray}
u(1+\alpha e^{-u^2})=u^\prime.\label{lens}
\end{eqnarray}
Here
\begin{eqnarray}
\alpha=\frac{c^2 r_{\rm e} {\rm DM_l}}{\pi a^2\nu^2}\left(\frac{d_{\rm sl}d_{\rm lo}}{d_{\rm so}}\right)=\frac{3430~{\rm DM_l}d_{\rm sl,kpc}}{\nu_{\rm GHz}^2a_{\rm AU}^2}\left(\frac{d_{\rm lo,Gpc}}{d_{\rm so,Gpc}}\right),\nonumber\\\label{alpha}
\end{eqnarray}
where $\nu_{\rm GHz}=\nu/\unit{GHz}$, $d_{\rm sl,kpc}=d_{\rm sl}/\unit{kpc}$, $d_{\rm lo,Gpc}=d_{\rm lo}/\unit{Gpc}$, $d_{\rm so,Gpc}=d_{\rm so}/\unit{Gpc}$, $a_{\rm AU}=a/\unit{AU}$, and ${\rm DM_l}$ has the units of $\unit{pc~cm^{-3}}$. For a given $u^\prime$, if $\alpha>\alpha_{\rm m}\equiv e^{3/2}/2$, Eq.(\ref{lens}) would have one to three solutions to $u$, which correspond to one to three sub-images of an FRB. 

Following \citet{cor17}, we consider a plasma lens with a dispersion depth ${\rm DM_l}=10~\unit{pc~cm^{-3}}$, and adopt $d_{\rm sl,kpc}=1$, $d_{\rm so,Gpc}\simeq d_{\rm lo,Gpc}=1$, $a_{\rm AU}=60$. The lens parameters give $\alpha=9.5\nu_{\rm GHz}^{-2}$. The focal frequency is $\nu_{\rm f}=\nu(\alpha/\alpha_{\rm m})^{1/2}\sim 2~\unit{GHz}$, which is comparable to the typical FRB observation frequencies. As pointed out by \citet{cor17}, for a certain gain threshold, e.g. $G>5$, most events have only one image or sub-image above the gain threshold. 
Some doublet cases occur at the same location and frequency but arrive at different times.
The number of triplet cases is small, and most of them are near the focal frequency $\nu_{\rm f}\sim2~\unit{GHz}$ with an arrival time less than $10~\unit{ms}$. 

For a given $u^\prime$, the DM perturbation occurs at a certain narrow frequency band where the gain is $G\gg1$, leading to the apparent DM deviating from ${\rm DM}(u)={\rm DM_l}\exp(-u^2)$ \citep{cor17}.
At another frequency, the DM perturbation could be neglected.
Due to the motion of the source or the observer relative to the lens, the DM and its perturbation can show an evolution. 
According to Eq.(\ref{up}), the effective transverse velocity is given by
\begin{eqnarray}
v_\perp^\prime=\frac{d_{\rm lo}}{d_{\rm so}}v_{\rm s, \perp}+\frac{d_{\rm sl}}{d_{\rm so}}v_{\rm o, \perp}.
\end{eqnarray}
In general, since $v_{\rm s, \perp}\sim v_{\rm o, \perp}$ and $d_{\rm sl}\ll d_{\rm so}$, one has $v_\perp^\prime\sim(d_{\rm lo}/d_{\rm so})v_{\rm s, \perp}\sim v_{\rm s, \perp}$. Here, we take $v_\perp^\prime\sim 200~\unit{km~s^{-1}}$ as a typical virial velocity in a galaxy. 
The characteristic time scale of a caustic crossing ($G>1$) is given by \citet{cor17}
\begin{eqnarray}
t_{\rm cau}\simeq\frac{a(\delta G/G)}{v_\perp^\prime G^2}&\simeq&21~\unit{day}\left(\frac{\delta G}{G}\right)\left(\frac{G}{5}\right)^{-2}\nonumber\\
&\times&\left(\frac{a}{60~\unit{AU}}\right)\left(\frac{v_\perp^\prime}{200~\unit{km~s^{-1}}}\right)^{-1}
\end{eqnarray}
where $\delta G$ is the corresponding change in gain. For the given typical parameters, one has $\delta{\rm DM}/{\rm DM_l}\sim0.1$ \citep{cor17}. Therefore, the DM variation during a caustic crossing may be 
\begin{eqnarray}
\left|\frac{d{\rm DM_{cau}}}{dt}\right|&\simeq&\frac{|\delta{\rm DM}|}{t_{\rm cau}}\simeq17~\unit{pc~cm^{-3}yr^{-1}}\left(\frac{{\rm DM_l}}{10~\unit{pc~cm^{-3}}}\right)\nonumber\\
&\times&\left(\frac{\delta G}{G}\right)^{-1}\left(\frac{G}{5}\right)^{2}\left(\frac{a}{60~\unit{AU}}\right)^{-1}\left(\frac{v_\perp^\prime}{200~\unit{km~s^{-1}}}\right)\nonumber\\
\end{eqnarray}
For the non-gained frequency or a long-term observation with $T_{\rm obs}\gg t_{\rm cau}$, the DM perturbation could be neglected, and the apparent DM only depends on the DM profile of the plasma lens. Here we simply consider a range of $u^\prime\sim 2-10$.
According to Eq.(\ref{lens}), one has $u\sim0.2$ at $u^\prime\sim2$ and $u\sim10$ at $u^\prime\sim10$ for $\alpha\sim9.5$. Thus, $\Delta{\rm DM}\sim {\rm DM}(u\sim0.2)-{\rm DM}(u\sim10)\sim9.6~\unit{pc~cm^{-3}}({\rm DM_l/10~\unit{pc~cm^{-3}}})$. The corresponding DM variation may be approximately estimated as
\begin{eqnarray}
\left|\frac{d{\rm DM}}{dt}\right|&\simeq&|\Delta{\rm DM}|\left(\frac{\delta u^\prime a}{v_\perp^\prime}\right)^{-1}\simeq0.8~\unit{pc~cm^{-3}yr^{-1}}\nonumber\\
&\times&\left(\frac{{\rm DM_l}}{10~\unit{pc~cm^{-3}}}\right)\left(\frac{a}{60~\unit{AU}}\right)^{-1}\left(\frac{v_\perp^\prime}{200~\unit{km~s^{-1}}}\right).\nonumber\\
\end{eqnarray}
We note that this value is much larger than the DM variation caused by the gas density fluctuation in the ISM (cf. Eq.(\ref{eq:DM-var-ISM})).
The reason is that plasma lensing is required to be dense and concentrated.

\section{Discussion and Conclusions}

The DM variations of Galactic pulsars are of the order $(10^{-5}-0.1)~\unit{pc~cm^{-3}~yr^{-1}}$ \citep[e.g.][]{isa77,ham85,lam16,jon17}, which originate from a variety of ways, with components that appear linear, periodic, or random. The long-term linear and periodic variation might result from changing distances between the pulsar and Earth, pulsar motion through the enveloping medium, accumulated effects along the line of sight,  etc. \citep[e.g.][]{lam16,jon17}. Compared with Galactic pulsars, FRB 121102 is of a cosmological origin, likely originating from a younger neutron star surrounded by a young SNR or PWN \citep[e.g.][]{mur16,pir16}.  It is possible that the FRB source can provide observable DM variations during long-term observations.

In this paper, we have discussed some astrophysical processes that may cause the DM variation of a repeating FRB source. The processes span from the cosmological scale to the local scale, including Hubble expansion, gravitational potential fluctuations and gas density fluctuations of LSS, gas density fluctuations of the ISM, SNR expansion, PWN evolution, HII region effect, and plasma lens. We find that the DM variations due to  large-scale effects are negligibly small, and only the local effects, such as SNR, PWN, HII region and plasma lens, can cause observable DM variations of FRBs.

In particular, different from previous works \citep[e.g.][]{kat16,met17,yan17,pir17,kat17} about the free-expansion SNRs with the negative DM variation\footnote{ \citet{pir16} discussed both the free-expansion and Sedov-Taylor phases and included the ionization changes, based on an assumption that the SNR mass is dominated by the ejecta.}, we find that a SNR can provide a positive DM variation during the Sedov-Taylor and snowplow deceleration phases, when it expands in a nearly neutral ambient medium. Although the DM variation is smaller than that in the free-expansion phase, the late phases can last a longer time. If the ambient medium is dense enough, e.g. $n\gtrsim100~\unit{cm^{-3}}$, where $n$ is the number density of the ambient medium, the DM variation in the Sedov-Taylor phase can be observable within $\sim 10~\unit{yr}$.

For a PWN, since the dense pair-wind plasma is close to a magnetized neutron star, one needs to consider the dispersion relation in a magnetized plasma. Four electromagnetic wave modes, i.e. O mode, X mode, R mode and L mode, are considered.
We find that only O-mode waves can show significant DM and DM variation. 
We also consider the DM variation from the thermal non-relativistic electron-positron pairs in the shocked region \citep{yan17}, which is much smaller than that of the relativistic pair wind.
Since the O-mode waves (with linear polarization) have a larger contribution to DM and DM variation of a PWN than the R-mode and L-mode waves (with circular polarization), a linear-polarization-dependent DM and DM variation may be useful to test the PWN hypothesis in the future observations.

An HII region might cause DM variation in two ways: 1. an expanding HII region associated with a new-born star; 2. the FRB source moving rapidly within the HII region. 
In the first case, the HII region is very young (age $\lesssim1000~\unit{yr}$) and the ionization process is dominant. The DM of the HII region increases due to the injection of the ionizing photons from the young star. 
However, the time window for this to happen may be very short within the astronomical context.
In the second case, the HII region has reached the Str\"omgren radius, and the DM variation results from the relative motion (including the line of sight and transverse motions) between the FRB source and HII region.

Plasma lens can cause multi-image bursts, which show different apparent DMs and different arrival times \citep{cle98,cor17}. 
For a lens with the DM depth of ${\rm DM_l}\sim10~\unit{pc~cm^{-3}}$ and the scale of $a\sim60~\unit{AU}$, the DM perturbation is $\delta{\rm DM}\sim1~\unit{pc~cm^{-3}}$ and the delay of the arrival time is $\lesssim10~\unit{ms}$ \citep{cor17}. In this work, we calculate the variation of DM and its perturbation, which are of the order of $\sim10~\unit{pc~cm^{-3}yr^{-1}}$ for the typical parameters. The corresponding evolution tendency depends on the direction of the transverse velocity of the source or the observer.

The recent observation of FRB 121102 showed that the DMs of the VLA bursts are slightly larger than those of the Arecibo bursts. 
Meanwhile, there is a weak correlation between the burst width and the apparent DM for VLA bursts \citep{law17}.
The burst-specific DM structure might be intrinsic to the emission mechanism or induced by the plasma lens effect \citep{law17,cor17}.
The effect of the burst intrinsic emission mechanism on the apparent DMs is complex, which depends on the poorly understood radiative mechanism of FRBs. For the plasma lens effect, the simultaneous detection of a burst of FRB 121102 from VLA and Arecibo requires a focal frequency of $\nu_{\rm f}\gtrsim$ 3.2 GHz and sufficient amplification in both bands \citep{law17,cor17}. If the intrinsic DM of FRB 121102 does not show secular variation over a long period of time as the current observations suggest, important constraints can be placed on the age of the FRB source and the existence of a SNR or PWN. If in the future, intrinsic variation of DM is detected from this source or other repeating FRBs to be detected in the future, important insights on the source physics can be gained by confronting the variation data with various mechanisms discussed in this paper.

\acknowledgments
We thank the anonymous referee for detailed suggestions that have allowed us to improve the presentation of this manuscript. We also thank Zhuo Li, Anthony Piro, Rui Luo, Yan Huang and Tian-Qi Huang for helpful comments and discussions.
This work is partially supported by Project funded by the Initiative Postdocs Supporting Program (No. BX201600003), the National Basic Research Program (973 Program) of China (No. 2014CB845800), and China Postdoctoral Science Foundation (No. 2016M600851). Y.-P.Y. is supported by a KIAA-CAS Fellowship.

\end{document}